\documentclass[draftcls,onecolumn,letter]{IEEEtran}

\usepackage{amsmath} % assumes amsmath package installed
\usepackage{amssymb}  % assumes amsmath package installed
\usepackage{mathrsfs}  % assumes amsmath package installed
\usepackage{arydshln}  % assumes amsmath package installed
\usepackage{color}
\usepackage[dvips]{graphicx}
\definecolor{gray}{gray}{.8}

\newcommand{\rev}[1]{{\color{black}#1}}

\newtheorem{remark}{Remark}
\newtheorem{theorem}{Theorem}

\newtheorem{proposition}{Proposition}

\newtheorem{problem}{Problem}

\newcommand{\R}{{\mathbb R}}

\newcommand{\C}{{\mathbb{C}}}

\newcommand{\tmk}{^{\mbox {\sf\scriptsize T}}}

\newcommand{\vmat}[2]{\left[
    \begin{array}{c}
      #1\\#2
    \end{array}\right]}
\newcommand{\dmat}[4]{\left[
    \begin{array}{cc}
      #1&#2\\
      #3&#4
    \end{array}\right]}

\newcommand{\hmat}[2]{\left[
    \begin{array}{cc}
      #1&#2
    \end{array}\right]}

\newcommand{\vmatthree}[3]{
  \left[
    \begin{array}{c}
      {#1}\\{#2}\\ \hdashline[2pt/2pt] {#3}
    \end{array}
  \right]}

\newcommand{\dmatthree}[9]{
  \left[
    \begin{array}{c;{2pt/2pt}cc}
      #1&#2&#3 \\ \hdashline[2pt/2pt]
      #4&#5&#6 \\
      #7&#8&#9
    \end{array}
  \right]}

\newcommand{\tr}{\mathop{\rm tr}\nolimits}
\newcommand{\diag}{\mathop{\rm diag}\nolimits}
\newcommand{\ssp}{{\mathcal S}}    %% state space symbol
\newcommand{\pr}{{\mathbb P}}   %% probablity
\newcommand{\ex}{{\mathbb E}}   %% expectation
\newcommand{\exb}[1]{{\mathbb E}\left[{#1}\right]}   %% expectation
\newcommand{\io}{{\mathscr A}}   %% infinitesimal operator
\newcommand{\half}{\frac{1}{2}}   %% 1/2
\newcommand{\dint}[2]{\displaystyle{\int _{#1}^{#2}}}
\newcommand{\rmd}{{\rm d}}
\newcommand{\xb}{\tilde x}
\newcommand{\onemk}{^{\mbox{\tiny (1)}}}
\newcommand{\twomk}{^{\mbox{\tiny (2)}}}
\newcommand{\threemk}{^{\mbox{\tiny (3)}}}
\newcommand{\fmk}{_{\mbox{\scriptsize f}}}
\newcommand{\dmk}{_{\mbox{\scriptsize d}}}
\newcommand{\domC}{{\sf C}}
\newcommand{\domD}{{\sf D}}

\newcommand{\im}{{\rm i}}
\begin{document}
\title{Control of Quantum Systems Despite Feedback Delay}
\author{Kenji~Kashima\thanks{K.~Kashima is with Graduate School of Information
    Science and Engineering, Tokyo Institute of Technology, Tokyo, 152-8552,
    JAPAN (e-mail: {\tt
    kashima@mei.titech.ac.jp}).  Mailing address: W8-1, 2-12-1, O-okayama,
    Meguro-ku, Tokyo, Japan.  Tel.~\& Fax.:
    +81-3-5734-2646.} %,~\IEEEmembership{Member,~IEEE}
    and~Naoki~Yamamoto\thanks{N. Yamamoto is with the Department of 
        Applied Physics and Physico-Informatics, Keio University, 
        Hiyoshi 3-14-1, Kohoku-ku, Yokohama 223-8522, Japan 
        (e-mail: {\tt yamamoto@appi.keio.ac.jp}).}}
\maketitle

\begin{abstract}
 Feedback control (based on the quantum continuous measurement) of 
 quantum systems inevitably suffers from estimation delays. 
 % Control systems taking these delays into account can be described by 
 % using a class of stochastic delay differential systems. 
 In this paper we  give a delay-dependent stability criterion for a 
 wide class of nonlinear stochastic systems including quantum spin 
 systems. We utilize a semi-algebraic problem 
 approach to incorporate the structure of {\em density matrices}. To show 
the  effectiveness of the result, we derive a globally stabilizing 
control law for a quantum spin-1/2 systems in the face of feedback delays. 
%
%  A quantum control scenario follows the formalism of continuous
%quantum measurement and quantum state estimation. In quantum feedback
%control implementations the state estimation is delayed, which is
%known to degrade the resulting control performance. As yet there are
%no theoretical means to take such estimation delays into
%consideration.  In this paper, we formulate control problems for
%general quantum systems taking feedback delays into explicit account.
%Then we derive a globally stabilizing control law for quantum spin
%systems.
\end{abstract}


% Will be ignored for other modes.
\IEEEpeerreviewmaketitle

\section{Introduction}\label{Section:introduction}

Quantum systems substantially differ from classical (i.e., non-quantum) 
systems in that state variables are represented by noncommutative 
operators acting on a Hilbert space; see e.g., \cite{sakurai}.  Such 
noncommutativity imposes some critical constraints on the structure of a 
quantum controller. This makes it difficult to analyze/synthesize 
feedback control systems for quantum systems. However, {\it quantum 
filtering theory} \cite{belavkin1,belavkin2,belavkin3,bouten2} has 
clarified that a number of quantum control problems can be formulated 
and solved within the framework of standard classical stochastic control 
theory \cite{bouten1,doherty,ramon, 
james,mirrahimi,thomsen,wiseman,yamamoto,ahn1,ahn2}.

A brief description of the filter-based approach to quantum control 
is as follows. 
The plant dynamics are given by a {\it quantum stochastic 
differential equation}, where the state is a noncommutative random 
variable \cite{hudson}. The dynamics are partially monitored by means 
of a continuous measurement that allows us to construct an 
estimator of the plant variables. The resulting filter is a 
classical stochastic differential equation called the 
{\it Belavkin equation} or {\it stochastic master equation}. 
Our objective is to synthesize an effective controller such that 
the filter shows a desirable behavior.

For this problem, two types of control law have been proposed. 
The first one is a simple proportional feedback of the output signal. 
%This control law was demonstrated to be successful for spin 
%squeezing control \cite{geremia,stockton}. 
%However, in general, this control law does not guarantee stability; 
%see the next section for detail. 
The second one is a feedback of the estimate of the plant variables, 
which we call the {\it filter-based controller}. 
A more detailed description of these two controllers will be given 
in the next section, but we here note that, for the implementation 
of the filter-based controller, a non-negligible computation time is 
required to process the estimation \cite{prog}. 
Therefore, from a practical point of view, a filter-based feedback 
controller should be considered taking the feedback delay into 
explicit account. 
For example, Steck {et al.} have numerically studied the issue of delay 
in the case of feedback stabilization of atomic motion \cite{steck}. 
However, to the authors' best knowledge, there have been no theoretical 
means to perform such investigations in the quantum case.

In this paper we study the effect of the delay in quantum systems with 
the full use of several techniques for analyzing stability of stochastic 
delay differential systems; see e.g., \cite{mao,mohammed,yue} and 
references therein. In particular, we focus on the control problem of a 
quantum spin system, which has also been studied in 
\cite{ramon,mirrahimi,yamamoto}. This system is very important, since it 
is one of the most basic components in quantum information processing 
\cite{nielsen}.

This paper is organized as follows. Section II reviews quantum filtering 
and control. In particular, we discuss delay in this feedback control 
scheme. Section III is the main part of this paper. Theorem 1 gives a 
delay-dependent stability criterion for a class of nonlinear stochastic  
systems including some quantum spin systems. The effectiveness of the 
result is then verified by deriving a stabilizing controller for the 
spin-$1/2$ particle case.

\vspace{5mm}\noindent{\bf Notation} \hspace{.5cm}
For $z\in \R^n$ and $M\in \R^{n\times n}$, $\|z\|^2_M := z\tmk M z$.
The subscript is omitted when $M$ is the identity matrix.  A function $F: 
\domD \to \R$ is said to be \emph{negative} (resp. \emph{positive}) in $\domD$ 
if $F(z) \le 0$ (resp. $F(z)\ge 0$) for any $z\in \domD$.
A subset $\domC$ in $\R^n$ is said to be \emph{semi-algebraic} if 
\[
 \domC := \{x\in \R^n: p_i(x)\le 0, \ i=1,2,\cdots, l\}
\]
with polynomials $p_i$.

Let $C_\domC^h$ be the set of $\domC$-valued uniformly continuous 
functions on $[-h,0]$. This is a Banach space equipped 
with $\|\xb\|_{C_\domC^h} := 
\sup_{\theta\in[-h,0]}\|\xb(\theta)\|_{\domC}$. Given a probability 
measure, the probability and expectation are denoted by $\pr$ and $\ex$. 
We say an event $\Omega$ occurs almost surely if 
$\pr\left\{\Omega \right\}=1$. 
If it exists, the {\em infinitesimal generator} of a function $V$ along 
a Markov process $\tilde x_t$ is denoted by $\io V$ i.e., 
$\displaystyle\io V(\tilde x) := \lim_{t\rightarrow 0} \frac{\ex^{\tilde 
x}[\tilde x_t] - \tilde x}{t}$ where $\ex^{\tilde x}$ represents the 
expectation with respect to paths which start at $\tilde x_0 = \tilde x$; 
see \cite{mohammed,mao,yue} for a formula.

\section{Control scheme based on Quantum filtering}

\subsection{Quantum filtering}

We here provide a brief summary of quantum filtering theory 
\cite{belavkin1,belavkin2,belavkin3}. 
For a more detailed description, see \cite{bouten2}.

In the framework of quantum filtering, a plant dynamics is 
described in a similar form to a general classical stochastic 
differential equation. 
For example, when using a {\it homodyne detector} \cite{gardiner}, 
a single state variable $X_t$ satisfies 
\begin{eqnarray}
& & \hspace*{-1em}
\label{plant-dynamics}
    \rmd X_t=f(X_t)\rmd t+g(X_t)\rmd W_t,
\nonumber \\ & & \hspace*{-0.7em}
    \rmd Y_t=(h(X_t)+h(X_t)^*)\rmd t+\rmd V_t,
\end{eqnarray}
where $f,g$, and $h$ are smooth functions with specific 
structures. 
However, unlike the classical case, the state variable $X_t$, the 
output $Y_t$, and the stochastic noises $W_t, V_t$ are 
{\it observables}, i.e., Hermitian operators that act on a certain 
Hilbert space ($*$ denotes the self-adjoint operation). 
Thus, in general they do not commute with each other. 
Note that any noncommutative random variables cannot take their 
realization values on a same probabilistic space. 
This implies that the classical stochastic control theory is not 
directly applicable, because we cannot define the conditional 
expectation $\pi(X_t):={\bf E}(X_t | {\cal Y}_t)$, and consequently
the optimal filter. 
Here, ${\cal Y}_t$ denotes the set of $Y_s~(0\leq s\leq t)$. 
Quantum filtering theory identifies systems free from these 
difficulties, i.e., systems satisfying the 
{\it nondemolition properties} $[Y_s, Y_t]=0~(\forall s,t)$ and 
$[X_t, Y_s]=0~(\forall s\leq t)$, where $[A,B]:=AB-BA$. 
Fortunately, in many important cases, especially in quantum 
optics, we can build such systems. 
The filter is then given by
\begin{eqnarray}
& & \hspace*{-1em}
\label{filter}
   \rmd\pi(X_t)=\pi(f(X_t,u_t))\rmd t
\nonumber \\ & & \hspace*{-1em}
   \mbox{}
     +\Big(\pi(X_t h(X_t)+h(X_t)^*X_t)
     -\pi(X_t)\pi(h(X_t)+h(X_t)^*)\Big)
\nonumber \\ & & \hspace*{3em}
   \mbox{}
     \times \big( \rmd Y_t-\pi(h(X_t)+h(X_t)^*)\rmd t \big).
\end{eqnarray}
Surprisingly, this is the same form as the classical filtering 
equation except the symmetrized terms. 
We now introduce a {\it density matrix} $\rho$; 
in a finite-dimensional case, it belongs to the convex set 
\begin{equation}
\label{state-space}
    {\cal S}:=\{\rho\in{\mathbb C}^{N\times N}~:~
                   \rho=\rho^*\geq 0, \tr\rho=1\}, 
\end{equation}
where $N$ is determined from the system. 
The statistics of the measurement results of an observable $X$ is 
completely characterized by $\rho$. 
For example, the $k$-th moment of the outcomes is 
given by $\tr(X^k\rho)$. 
Thus the conditional expectation $\pi(X_t)$ should also be 
represented in terms of a time-dependent density matrix $\rho_t$ as 
$\pi(X_t)=\tr(X\rho_t)$, which together with \eqref{filter} 
leads to the time-evolution of $\rho_t$. 
In particular, when the homodyne detection scheme is used, 
the most simple form of it is given by the following 
{\it stochastic master equation}: 
\begin{eqnarray}
\label{general-SME}
& & \hspace*{-1.2em}
    \rmd \rho_t
     ={\cal L}^*(\rho_t, u_t)\rmd t
         +\Big(L\rho_t+\rho_tL^*-\tr(L\rho_t+\rho_tL^*)\rho_t\Big)
\nonumber \\ & & \hspace*{9em}
   \mbox{}
     \times \big( \rmd Y_t-\tr(L\rho_t+\rho_t L^*)\rmd t \big), 
\nonumber \\ & & \hspace*{-1em}
    {\cal L}^*(\rho,u)
     :=\im[H,\rho]+L\rho L^*-\half L^*L\rho-\half \rho L^*L.
\end{eqnarray}
Here, $H$ is an observable called {\it Hamiltonian}, 
representing the energy of the system. 
The measurement operator $L$ determines how the system interacts 
with the measurement apparatus (e.g. a laser field; see Figure 1).

\subsection{Implementation of filter-based controller}

In a typical situation, the Hamiltonian term is a function of the 
control input $u_t$; $H=H(u)$. Our goal is to design $u_t$ such that the 
filter of Eq.~\eqref{general-SME} has a desirable behavior. 
Note that, as in the classical case, the last term 
$\rmd w_t:=\rmd Y_t-\tr(L\rho_t+\rho_t L^*)\rmd t$ 
is a classical Wiener increment. 
This implies that Eq. \eqref{general-SME} is a classical 
stochastic differential equation to which several techniques 
developed in control theory can be applied.

The proportional output feedback controller 
$u_t=k\rmd Y_t/\rmd t$ ($k\in{\mathbb R}$ is the gain) is often 
considered \cite{ahn2,thomsen,wiseman} and was implemented in the 
experimental setup of spin-squeezing control \cite{geremia,stockton}.
%However, with this control input, the resulting filter equation 
%(\ref{general-SME}) is generally not stable at the target density matrix. 
On the other hand, note that we can compute $\rho_t$ by using the 
past output sequence $\{Y_s\}_{s\le t}$ by Eq.~(\ref{atomicSME}). 
If it is possible to perform this computation on-line, we can implement 
controller of the state feedback form $u_t = u(\rho_t)$, i.e., 
the filter-based controller. 
With this control the target state is limited to the {\it eigenstates} 
of the measurement operator $L$ unlike the proportional feedback 
case where the target can be to some extent changed flexibly 
\cite{wang,ticozzi}, but we can instead take much wider variety 
of designing methods of the filter-based controller. 
In fact, it has been proven that the Lyapunov theory was successfully 
employed to show the global stability of the filter for some systems 
\cite{ramon,mirrahimi,yamamoto,altafini}. 
Moreover, it is known that the optimal controller for a general type of 
quantum optimal control problem is given by a filter-based controller. 
This is known as the {\it separation theorem} \cite{bouten3}.

However, in general, the time required to compute $\rho_{t}$ is not 
negligible compared to the time-constants associated with the dynamics 
of a nano-mechanical system. In other words, from a practical point of 
view, $\rho_t$ cannot be used to determine $u_t$. In view of this we 
should consider the delayed feedback control input 
$u_t=u(\rho_{t-\tau})$, where $\tau>0$ denotes the delay length. Note 
that this formulation is able to handle further delays, for example 
input delays. Such input delays occur because the control input $u_{t}$ 
must be physically implemented by means of actuators. The purpose of 
this paper is to propose a rigorous methodology for analyzing the 
behavior of quantum control systems in the face of feedback delay.

\section{STABILIZATION OF QUANTUM SPIN SYSTEMS IN THE FACE OF DELAY}

\subsection{The physical model and control problem}

In this section we consider a cold atomic ensemble trapped in 
an optical cavity \cite{geremia,ramon,mirrahimi,thomsen,yamamoto}, 
as depicted in Figure~1. 
The total angular momentum operator $F_i$ of the atom around the 
$i$-axis ($i=y,z$) is given by 
\begin{eqnarray*}
F_y&:=&\frac{\rm i}{2}
\left[
\begin{array}{ccccc}
0&c_1&&&\\
-c_1&0&c_2&&\\
&\ddots&\ddots&\ddots&\\
&&-c_{N-2}&0&c_{N-1}\\
&&&-c_{N-1}&0
\end{array}
\right],\\
c_m&:=&\sqrt{(N-m)m},\quad m=1,2,\cdots, N-1,\\
F_z&:=&\frac{1}{2} \diag\{ N-1,N-3, \\
	&& \hspace{2cm}\cdots, -(N-3), -(N-1)\}, 
\end{eqnarray*}
%
%\[
%    F_y=\half\left[ \begin{array}{cc}
%          0 & \im \\
%          -\im & 0 \\
%        \end{array} \right],~~
%    F_z=\half\left[ \begin{array}{cc}
%          1 & 0 \\
%          0 & -1 \\
%        \end{array} \right].
%\]
where $N-1$ represents the number of atoms. 
The system interacts with a laser field oriented along the $z$-axis at a 
homodyne-type photo detector, which implies $L=F_z$. The system also 
interacts with an external magnetic field, which is oriented along the 
$y$-axis, $H(u_t)=u_t F_y$, where the control input $u_t$ corresponds to 
the magnetic field strength, which can be modified in time. As a result, 
the controlled filter equation (\ref{general-SME}) becomes
\begin{eqnarray}
\label{atomicSME}
& & \hspace*{-1em}
    \rmd\rho_t=\im[u_t F_y,\rho_t]\rmd t-\half [F_z, [F_z, \rho_t]]\rmd t
\nonumber \\ & & \hspace*{3em}
    \mbox{}
        +\sqrt{\eta}\big(F_z\rho_t
            +\rho_tF_z-2\tr(F_z\rho_t)\rho_t\big)\rmd w_t,
\end{eqnarray}
where $\eta\in(0,1]$ represents the measurement efficiency. 
Note that the Wiener process $w_t$ contains the measurement data $y_t$.

Our goal is to design a feedback control law $u_t=u(\rho_{t-\tau})$ that 
achieves the deterministic convergence of $\rho_t$ to a prescribed 
target state. This problem was solved in \cite{ramon,mirrahimi,yamamoto}, 
for the case  of no delay. Note that controlled filter equation 
(\ref{atomicSME}) shows a significant dependence on the delay, 
through the input $u_t=u(\rho_{t-\tau})$. 
Therefore the control problem is much more difficult than the 
previous one.

\begin{figure}
\centering
\includegraphics[width=2.8in]{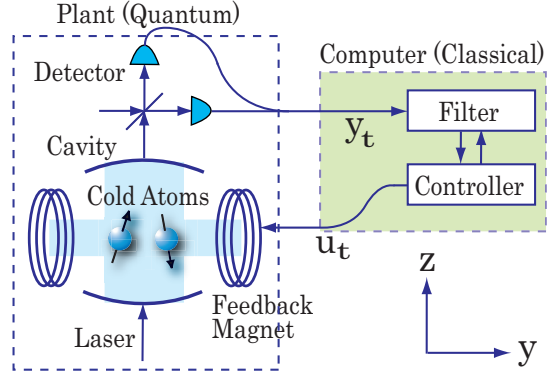}
\caption{
Quantum plant and classical controller.
The filter needs a finite time $\tau>0$ to compute the control input
$u_t=u(\rho_{t-\tau})$.
}
\end{figure}

\subsection{Delay-dependent stability criteria}

The system of Eq. (\ref{atomicSME}) is described by $\rho_t\in 
\C^{N\times N}$. By concatenating the real and imaginary part of
all elements of $\rho_t$ into a column vector, we can rewrite 
Eq.~(\ref{atomicSME}) as a $\R^n$-valued nonlinear stochastic delay 
system. It is important to note that the resulting system has the following 
features: 
\begin{itemize}
 \item The drift and diffusion terms are polynomials in the state variable. 
 \item The bounded semi-algebraic set determined by $\ssp$ is positively 
       invariant; see also \cite[Proposition 1]{ramon}.
 \item The control input, which
       possibly suffers from delays, is applied only to the drift term. 
\end{itemize}
We here do not limit our attention to the specifically  
structured dynamics of Eq.~(\ref{atomicSME}), but rather consider a 
wide class of nonlinear stochastic systems with the above properties. 
%In view of this, we analyze the stability of a class of stochastic delay 
%system having these properties. 
A delay-dependent stability 
criterion is given in Theorem \ref{maintheorem}\footnote{Throughout this section, the 
symbols $\tilde x_{\bullet}$ (resp. $x_\bullet$) are used to 
represent functions (resp. vectors). These symbols with (resp. without) 
the time index denote the solution to Eq. (\ref{sdde}) (resp. any 
functions or vectors).}.
\begin{theorem}\label{maintheorem}
  Let $f(\cdot, \cdot): \R^n \times \R^n \to \R^n$, $g(\cdot):  \R^n\to 
 \R^n$, be polynomials and $\domC$ a   
 bounded semi-algebraic set in $\R^n$ such that for any initial 
 condition $\tilde x_{\rm i} \in C_{\domC}^\tau$ the 
 solution to the delay differential stochastic equation
  \begin{eqnarray}\label{sdde}
    \rmd x_t & = & f(x_t, x_{t-\tau}) \rmd t + g(x_t) \rmd w_t \\
    x_\theta & = & \tilde x_{\rm i}(\theta) \in \domC,\quad \theta \in [-\tau,0]
    \label{sdde-ini}
  \end{eqnarray}
 does not exit $\domC$ \rev{almost surely}. Suppose there exist a 
 polynomial $V_*(\cdot)$ which is positive in $\domC$, 
 $n$-variable polynomials $V_i\ (i=0,1)$, $S\in \R^{2n\times n}$, and 
 positive-definite matrices $R,\ T\in \R^{n\times n}$  such that $\Upsilon$ defined below is negative in $\domC \times \domC \times 
 \R^{2n}$: 
  \begin{equation}\label{F}\begin{array}{l}
      F (x,x\dmk) := \left( \frac{\partial V_0(x)}{\partial x} \right)\tmk
      f(x,x\dmk) \\
      \hspace{6mm}+ \half\ g(x)\tmk \frac{\partial}{\partial x}
      \left( \frac{\partial V_0(x)}{\partial x} \right)\tmk g(x)\vspace{2mm}\\
      \vspace{2mm}\hspace{6mm}+V_1(x) - V_1(x\dmk)
      + V_*(x) + \tau \|g(x)\|_T^2\\
      \hspace{6mm}+ 2 \hmat{x\tmk}{x\dmk\tmk} S (x-x\dmk)
      + \tau \|f(x,x\dmk)\|_R ^2 \vspace{2mm}\\
      \Upsilon(x,x\dmk,y):=F(x,x\dmk) \vspace{2mm}\\
      \hspace{6mm}+ \vmatthree{x}{x\dmk}{y}\tmk
      \dmatthree{0}{S}{\tau S}{S\tmk}{-T}{0}{\tau S \tmk}{0}{-\tau R}
      \vmatthree{x}{x\dmk}{y}.
    \end{array}
  \end{equation}
Then, \rev{$V_*(x_t)$ converges to $0$ almost surely}
for any initial condition $\tilde x_{\rm i}\in C_{\domC}^\tau$.
\end{theorem}

Suppose that $V_*(x)$ represents a distance between $x$ and a given 
target state. Then, this theorem states that $x_t$ converges to the 
target state if a {\it semi-algebraic problem} is feasible; see 
also Subsection III.C. 
Semi-algebraic problems are in general NP-hard. However, if the degrees 
of polynomials have been decided, {\it sums of squares} (SOS) relaxation 
enables us to solve the problem efficiently \cite{parillo1,parillo2}. In 
the numerical example in the next subsection, we utilized MATLAB 
SOSTOOLS \cite{sostools,sedumi}.

\begin{remark}
 In Theorem \ref{maintheorem}, $\Upsilon$ is required to be negative  
 only in  $\domC \times \domC \times \R^{2n}$, not globally (i.e., in 
 $\R^{4n}$). This is the reason why Theorem 1 can incorporate the 
 structure of density matrices which is useful for reducing the 
 conservativeness. Similar criteria for  some modified problem 
 formulations (i.e., time-varying delay or  delay-independent stability) 
 can be obtained straightforwardly. 
 \hspace*{\fill}~\QED
\end{remark}

We prove Theorem \ref{maintheorem} by using the following
Lyapunov-Krasovskii type argument:
\begin{proposition}
  Let $x_t$ be the solution of the stochastic delay differential equations 
 (\ref{sdde}) and (\ref{sdde-ini}). Define
  \[
    \xb_t(\theta) := x_{t+\theta},\ \theta \in [-2\tau,0]
  \]
  for $t\ge \tau$. Suppose that there exists a
  positive functional $V$ defined in $C_\domC^{2\tau}$ such that 
  \begin{equation}\label{Lyap2}
    \exb{ \io V(\xb_t) + V_*(x_t) } \le 0
  \end{equation}
  for any $t \ge \tau$. Then, $V_*(x_t)$ converges to $0$ in the
  same sense as in Theorem \ref{maintheorem}.
\end{proposition}

\begin{proof}\rev{
  Recall that $x_t$ evolves only in the bounded domain $\domC$. Hence Fubini's 
theorem yields 
\[
\exb{\int_\tau ^t \io V(\xb_s) \rmd s} = \int_\tau ^t \exb{\io V(\xb_s)} \rmd s. 
\]
By combining this equality, Eq. (\ref{Lyap2}), and Dynkin's formula \cite{mohammed,kushnerD}, we 
obtain  
\begin{eqnarray*}
 \exb{V(\tilde x_t)} -V(\tilde x_\tau) &=& \exb{\int_\tau^t \io V(\xb_s) \rmd s} \\
  &\le& -\int_\tau ^t \exb{ V_*(x_t)} \rmd s \le 0.
\end{eqnarray*}
 Therefore we conclude that $V(\xb_t)$ is a 
 nonnegative super-martingale. The remainder of the proof is the same as the 
 standard Lyapunov-Krasovsii argument; see e.g. Theorems 6.1 and 6.2 in 
 \cite{kushnerD} and their proofs. This completes the proof.
}
\end{proof}

Now we are ready to prove Theorem 1. 

\begin{thmproof}
It suffices to show that $V$
\begin{equation}\label{Lyap1}
\begin{array}{l}
  V ( \xb ):= V_0( \xb(0) )
  + \dint{-\tau}{0} V_1( \xb(\theta) ) \rmd \theta \\
  \hspace{.0cm}+ \dint{-\tau}{0} \dint{v}{0}
  \left\{\| f(\xb(\theta),\xb(-\tau+\theta))\|_R^2
    + \|g(\xb(\theta))\|_T^2 \right\} \rmd \theta \rmd v
\end{array}
\end{equation}
satisfies the assumptions made in Proposition 1. 

 The polynomials $V_i\ (i=0,1)$ are bounded from below on $\domC$ due to the 
 continuity of polynomials and the boundedness of the domain. 
 Note that adding any constant to $V_i$ does not affect $\Upsilon$. 
 Therefore, without loss 
 of generality we can assume that $V$ is positive.

A direct computation yields 
\[
\begin{array}{l}
  0 \le \tau e^{\sf T} X e - \dint{-\tau}{0} e^{\sf T} X e \rmd s \\
  0 = (2-2)\cdot e^{\sf T} S \left\{ \xb(0) - \xb(-\tau) -\dint{-\tau}{0}
    \underline f(s) \rmd s \right\}\\
  \hspace{3mm}\le 2 e ^{\sf T} S (\xb(0) - \xb(-\tau))
  - \dint{-\tau}{0} 2e^{\sf T} S \underline f(s) \rmd s \\
  \hspace{6mm}+ e^{\sf T} S T^{-1}S^{\sf T} e
  +\left\| \xb(0)
    - \xb(-\tau)-\dint{-\tau}{0} \underline f(s) \rmd s \right\|_{T}^2
\end{array}
\]
where $e:= \hmat{\xb(0)\tmk}{\xb(-\tau)\tmk}\tmk$,
$\underline f(s) := f(\xb(s),\xb(-\tau+s))$, and $X := S
R^{-1}S^{\sf T} \ge 0$. Combining these inequalities and
\[
\begin{array}{l}
  \io V( \xb ) =  \left( \left. \frac{\partial V_0(x)}{\partial x}\right|_{\xb(0)} \right)\tmk f(\xb(0),\xb(-\tau)) \\
				\hspace{6mm}+ \half\ g(\xb(0))\tmk \left.\frac{\partial}{\partial x}
				\left( \frac{\partial V_0(x)}{\partial x} \right)\tmk \right|_{\xb(0)} g(\xb(0))\vspace{2mm}\\
				\vspace{2mm}\hspace{6mm}+V_1(\xb(0))
				- V_1(\xb(-\tau)) + \tau ( \|\underline f(0)\|_R^2 + \|g(\xb(0))\|_T^2 ) \vspace{2mm}  \\
  \hspace{6mm} - \dint{-\tau}{0} \left\{ \|\underline f(s)\|_R^2 + \|g(
\xb(s))\|_T^2 \right\} \rmd s,
\end{array}
\]
we obtain
\[
		\io V(  \xb ) + V_*(\xb(0))\le \tilde \Upsilon(\xb(0),\xb(-\tau)) -G_1(\xb) - G_2(\xb)
\]
with
\begin{eqnarray*}
	 \tilde \Upsilon(x,x\dmk) & := & F(x,x\dmk) + \left\|\hmat{x\tmk}{x\dmk\tmk}\tmk\right\|_{\tau X+S {T}^{-1}S\tmk } ^2\\
	G_1(\xb) & := & \int _{-\tau} ^0 \left\| \hmat{e\tmk}{\underline f(s)\tmk}\tmk\right\|_\Xi^2
	 \rmd s \ge 0\\
	G_2(\xb) & := & \int _{-\tau} ^0 \|g(\xb(s))\|_T^2 \rmd s \\
	& & - \left\| \xb(0) - \xb(-\tau) -\int _{-\tau} ^0 \underline f(s) \rmd s \right\|_{T}^2,\\
	\Xi &:=& \dmat{X}{S}{S\tmk}{R} \ge 0.
\end{eqnarray*}

Let us take the expectation after substituting $\xb = \xb_t$. We can show
\[
  \exb{G_2(\xb_t)} =
%  \exb{ \int _{-\tau} ^0 \|\xb_t(s)\|_S^2 - \| g(\xb_t(s))\|_{T}^2 \rmd s }
   0
\]
by using the It\^o isometry. We thus have
\[
  \exb{ \io V(\xb_t) + V_*(x_t) } \le
  \exb{ \tilde \Upsilon(x_t,x_{t-\tau}) } .
\]
Finally, by the assumption on $\Upsilon$ and defining
\[
 \bar y := \vmat{T^{-1}}{R^{-1}} S\tmk \vmat{x_t}{x_{t-\tau}}\in \R^{2n},
\]
we obtain
\[
 \tilde \Upsilon(x_t, x_{t-\tau} ) = 
 \Upsilon(x_t,x_{t-\tau}, \bar y ) \le 0.
\]
Therefore Eq.~(\ref{Lyap2}) follows. This completes the proof.
\end{thmproof}

\subsection{Numerical example: Control of a spin-1/2 system}
\label{sec:example}

This subsection focuses on a {\it spin-1/2} model such that the
system is composed of only a single particle. In this case, the density 
matrix $\rho_t$ is in $\C^{2\times 2}$. The filter equation 
(\ref{atomicSME}) without the input (i.e., $u_t=0$) shows the following 
probabilistic convergence: 
\[
    \rho_t\rightarrow\rho_{\uparrow}
    :=\left[ \begin{array}{cc}
          1 & 0 \\
          0 & 0 \\
     \end{array} \right]~~
    \mbox{or}~~
    \rho_t\rightarrow\rho_{\downarrow}
    :=\left[ \begin{array}{cc}
          0 & 0 \\
          0 & 1 \\
     \end{array} \right].
\]
This phenomenon is known as {\it quantum state reduction} \cite{adler}.
Here $\rho_{\uparrow}$ (resp. $\rho_{\downarrow}$) denotes the 
eigenstate (of $L=F_z$) for which the monitored spin state of the atom is 
deterministically up (resp. down). Note that when $u_t=0$, these two matrices are the only equilibrium points of Eq.~(\ref{atomicSME}). Our goal is to design a 
feedback control law $u_t=u(\rho_{t-\tau})$ that achieves the 
deterministic convergence of $\rho_t$ to the prescribed target 
$\rho\fmk$, which is either $\rho_{\downarrow}$ or $\rho_{\uparrow}$, as 
we choose.   

It is shown in \cite{ramon} that the control input $u_t = u(\rho_t)$ with
\begin{equation}\label{eqn:u}
    u(\rho) := k_1(1-\tr(\rho\rho\fmk)) + k_2 \tr(\im [F_y,\rho]\rho\fmk)
\end{equation}
achieves the control objective $\rho_t \rightarrow \rho\fmk$ when both $k_1$ 
and $k_2$ are chosen appropriately\footnote{
The interpretation of this control law is as follows. The second term 
(containing $k_2 >0$) \emph{locally} stabilizes $\rho\fmk$. 
Unfortunately, both $\rho_{\uparrow}$ and $\rho_{\downarrow}$ are  
equilibria of the closed-loop  
system. Hence, when $\rho_t$ is close to the eigenstate that is not the 
regulation point, $\rho_t$ must be prevented from converging to it. This 
is done by the first term. See \cite{kashima} for a discussion on the effect of delays when a switching control law is employed instead.}.  
In this subsection, we derive a sufficient condition for this control 
law to globally stabilize the spin-1/2 system in the face of 
feedback delay.  

Let us rewrite Eq. (\ref{atomicSME}) in terms of the regulation error
\[
  \dmat{x\onemk_t}{x\twomk_t+\im x\threemk_t}
  {x\twomk_t - \im x\threemk_t}{-x\onemk_t} := 
  \left\{
    \begin{array}{cl}
      \rho\fmk - \rho_t, & {\rm if\ } \rho\fmk = \rho_\uparrow\\
      \rho_t - \rho\fmk, & {\rm if\ } \rho\fmk = \rho_\downarrow.
    \end{array}
  \right.
\]
It can easily be verified that $\rho_t \rightarrow 
\rho\fmk$ is equivalent to $x_t := \hmat{x_t\onemk}{x_t\twomk}\tmk 
\rightarrow 0$. When we apply the control input $u_t = u(\rho_{t-\tau})$ 
with $u(\cdot)$ given by Eq. (\ref{eqn:u}), the dynamics of $x_t$ are 
independent of $x_t\threemk$ and are given by Eq. (\ref{sdde}) with 
\begin{eqnarray*}
  f(x,x\dmk) & := & \left[
    \begin{array}{c}
      - k x\dmk x \twomk\label{sdde3} \\
      \ k x\dmk \left( x \onemk - \frac{1}{2}\right) -\frac{1}{2} x \twomk
    \end{array} 	
  \right],\\
  g(x) & := & \sqrt{\eta}
  \vmat{2x \onemk (x \onemk - 1)}{(2 x \onemk - 1) x \twomk }, 
  \label{sdde4} \\
  k &:=& \hmat{k_1}{k_2}.
\end{eqnarray*}
Note that $\rho_t \ge 0$ means $x_t$ is in the 
circular domain $\domC$ 
\[
  \domC : = \left\{ \vmat{x\onemk}{x\twomk} \in \R^2:
      \Psi(x):=x\onemk(x\onemk-1)+{x\twomk}^2 \le 0 \right\}.
\]
It can be verified that, independently of $u_t$, the solution of Eq. (\ref{sdde}) does not 
exit $\domC$ almost surely. In summary, according to Theorem~1, if the 
following SOS decomposition problem has a solution, then the control 
objective is achieved:
\begin{problem}\label{prob:SOS}
 With the definitions above, let $V_*(x) := \|x\|^2$. Then, find 
$S\in   
\R^{4 \times 2}$, positive-definite matrices $R,\ T\in \R^{2\times 
 2}$, and polynomials $V_i\ (i=0,1)$, $h,\ h\dmk$ 
such that  
\begin{eqnarray*}
 &&-\Upsilon(x,x\dmk,y) - h(x,x\dmk,y)\Psi(x) - h\dmk(x,x\dmk,y)\Psi(x\dmk),\\ 
 &&h(x,x\dmk,y), \\
 &&h\dmk(x,x\dmk,y) 
\end{eqnarray*}
are the sum of squares of polynomials in $x,x\dmk \in \R^2$ and $y\in \R^4$.
\end{problem}
%the following two
%statements are equivalent\footnote{The dynamics of $x\threemk$ can 
%be ignored. Recall that $-w_t$ is also a standard wiener process.}:
%\begin{itemize}
%\item With $u(\cdot)$ given by (\ref{eqn:u}), $u_t = u(\rho_{t-\tau})$
%      stabilizes the target state $\rho\fmk$.
%\item The origin is stable in Problem 1 with the
%circular domain $\domC$ given by
%\[
%  \domC : = \left\{ \vmat{x\onemk}{x\twomk} \in \R^2:
%      \Psi(x):=x\onemk(x\onemk-1)+{x\twomk}^2 \le 0 \right\}
%\]
%and
%\begin{eqnarray*}
%  f(x,x\dmk) & := & \left[
%    \begin{array}{c}
%      - k x\dmk x \twomk\label{sdde3} \\
%      \ k x\dmk \left( x \onemk - \frac{1}{2}\right) -\frac{1}{2} x \twomk
%    \end{array} 	
%  \right],\\
%  g(x) & := & \sqrt{\eta}
%  \vmat{2x \onemk (x \onemk - 1)}{(2 x \onemk - 1) x \twomk }, 
%  \label{sdde4} \\
%  k &:=& \hmat{k_1}{k_2},\\ 
%  x &:=& \hmat{x\onemk}{x\twomk}\tmk. 
%\end{eqnarray*}
%\end{itemize}

We provide a numerical example to illustrate the effectiveness of
Theorem \ref{maintheorem}.  Decision polynomials are restricted
to quadratic functions. 
Let $k_1 = 1.0$ and $k_2=4.0$ which gives the control law whose stabilizing 
effect for the delay-free case was examined in \cite[subsection 
IV.G]{ramon}. Other parameters are chosen to be $\eta=0.9$ and $\tau = 0.3$. 
In this case, Problem \ref{prob:SOS} has a solution; that is, the target  
state in the controlled system is shown to be stable. 
It took 3.01 seconds to check the feasibility of Problem~1 using a
computer with a Pentium 4 3.2GHz processor and 2 GB memory. 

\begin{figure}[t]
 \setlength{\unitlength}{0.9mm}
 \begin{center}
 \begin{picture}(85,70)
% 	    \put(6,5){\includegraphics[width=7cm,height=5cm]{Eps/time_resp.eps}}
 	    \put(6,5){\includegraphics[width=7cm,height=5cm]{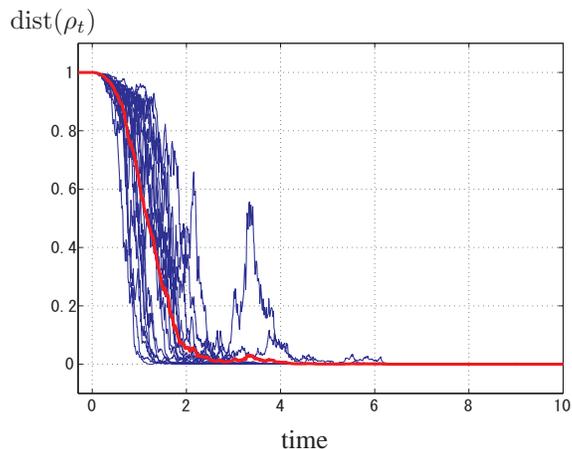}}
 		\put(40,0){time}
 		\put(0,62){${\rm dist}(\rho_t)$}

 \end{picture}
 \caption{Time responses of sample paths (thin blue lines) and their 
  average process (thick red line).} 
\label{fig:time}
\end{center}
 \end{figure}

By setting the target state $\rho\fmk := 
\rho_\uparrow$, we performed a numerical simulation. 
Time responses of the function
\[
{\rm dist}(\rho) := 1-\tr(\rho\rho\fmk): \ssp \rightarrow [0,1]
\]
are shown in Figure \ref{fig:time} (30 sample
paths and their average).  This function gives the distance from the
target state, i.e., ${\rm dist}(\rho) = 0$ (resp. ${\rm dist}(\rho) =
1$) if and only if $\rho = \rho\fmk$ (resp. $\rho = \rho_{\downarrow}$). 
The initial state is given by $\rho_t\equiv\rho_{\downarrow}$ for 
$-\tau\le t\le 0$. From Figure~\ref{fig:time} it can be seen that 
stability is achieved.  

\begin{remark}
In principle, the numerical approach introduced in this subsection is 
 applicable to the stability analysis of the general 
 multi-spin system despite time-delays. The computational complexity 
 grows quickly with the dimension. Very high dimensional problems are 
 therefore computationally intractable. On 
 the other hand, there exist some {\it  
 analytical} results for the $N$-dimensional delay-free case 
 \cite{mirrahimi,altafini}. The authors are currently investigating 
 computational approaches which combine the aforementioned numerical and 
 analytical methods, in order to overcome this computational issue.
 \hspace*{\fill}~\QED
\end{remark}

% \begin{figure}[t]
% \setlength{\unitlength}{0.9mm}
% \begin{center}
% \begin{picture}(85,70)
%% 	    \put(0,0){\includegraphics[width=7cm,height=6.5cm]{Eps/max_tau.eps}}
%        \put(0,0){\includegraphics[width=7cm,height=6.5cm]{max_tau.eps}}
% 		\put(60,0){$|k_1|$}
% 		\put(17,4){$k_2$}
%%   		\put(45,18){\scriptsize$0.05$}
%%   		\put(40,28){\scriptsize$0.10$}
%%   		\put(23,28){\scriptsize$0.15$}
%%   		\put(15,25){\scriptsize$0.18$}
%
%
% \end{picture}
% \caption{Gain $(k_1,k_2)$ versus the maximal delay length 
%   for which global stability is guaranteed}
% \label{fig:maxdelay}
% \end{center}
% \end{figure}

\section{CONCLUSION}

From a practical point of view, filter-based quantum control problems 
should be formulated taking feedback delay into explicit account. A 
delay-dependent stability criterion was derived for a class of nonlinear 
stochastic systems including some quantum spin control systems. A 
semi-algebraic approach was shown to be useful for incorporating 
the structure of density matrices.  

Theorem 1 was motivated by quantum spin control systems. Theorem 
\ref{maintheorem} can deal with any stochastic delay system having the 
three properties listed above it. Many finite-dimensional quantum 
systems satisfy these properties. Hence Theorem \ref{maintheorem} is 
applicable to a wide class of finite-dimensional quantum systems. 

This paper is a first attempt to analyze quantum systems which 
suffer from feedback delays. Hence, many important and interesting 
problems are left unsolved. The research topic mentioned in Remark 2 is 
one of them.

\bibliographystyle{IEEEtran}
%\bibliography{IEEEabrv,OTHERabrv}

\end{document}